\documentclass[preprint,11pt]{article}
\usepackage{amssymb}
\usepackage{caption}
\usepackage{breqn}
\usepackage{graphicx}

\usepackage{mathtools}
\DeclarePairedDelimiter\bra{\langle}{\rvert}
\DeclarePairedDelimiter\ket{\lvert}{\rangle}
\DeclarePairedDelimiterX\braket[2]{\langle}{\rangle}{#1\,\delimsize\vert\,\mathopen{}#2}
\usepackage{subcaption}
\usepackage{array,multirow}

\usepackage{cite}

\begin{document}

\textbf{Pomeron evolution, entanglement entropy and Abramovskii-Gribov-Kancheli cutting rules
}
\begin{center}
\small
 Mustapha Ouchen     and Alex Prygarin  
\\
  Department of Physics, Ariel University, Ariel 40700, Israel 
\end{center}
\normalsize

\normalsize

\begin{abstract}
  We use Pomeron evolution equation in zero transverse dimension  based on the Abramovskii-Gribov-Kancheli~(AGK)  cutting rules to calculate the von Neumann entropy and $q$-moments that used as an experimental test for the Koba-Nielsen-Olesen~(KNO) scaling. In order to avoid the negative probabilities that emerge   from   the negative AGK weights in the Minkowski space, we reformulate 
  the Pomeron evolution in the Euclidean space. The resulting positive definite probabilities for cut and uncut Pomerons are used in calculating the  $q$-moments. 
  The comparison to the experimental data shows that our AGK based model successfully describes  $q$-moments dependence on the mean multiplicity  without any adjustable parameter in the experimental data of the  $\mathtt{p-p}$ collisions by  $\mathtt{ALICE}$ Collaboration.

\end{abstract}

\section{Introduction}
In this paper   we apply the analysis done by  Levin and one of the authors~\cite{Levin:2008cgz} using the  Abramovskii-Gribov-Kancheli~\cite{AGK} rules and derive the von Neumann entropy 
for studying the entanglement in the high energy scattering.
 This continues the project started by Levin and Kharzeev~\cite{Kharzeev:2017qzs} which initiated a lot of activity related to that 
 topic~\cite{Kharzeev:2021yyf,Zhang:2021hra,Florio:2023dke,Hentschinski:2023izh,Hentschinski:2023nhy,Kutak:2023cwg,Kutak:2024pfy,Hatta:2024lbw,Levin:2024wtl,Kutak:2025hzo,Moriggi:2024tbr,Caputa:2024xkp,Bhattacharya:2024sno,Liu:2022bru}, other QCD related studies~\cite{Ramos:2025tge,Mamo:2025hur,Qi:2025onf,Ramos:2025uqn,Moriggi:2024tiz,Hutson:2024xsv,Dosch:2023bxj,Dumitru:2023qee,Duan:2023zls,Ramos:2022gia,Afik:2022kwm,Ramos:2020kaj,Peschanski:2019yah,Armesto:2019mna,Kovner:2018rbf} and beyond\cite{Duan:2020jkz,Baty:2021ugw,Florio:2021xvj,Grieninger:2023ehb,Miller:2023snw,Gursoy:2023hge,Barata:2023jgd,Grieninger:2023ufa,Khor:2023xar,Florio:2023mzk,Barr:2024djo,Meurice:2024fya}.

 Our goal is to build a simple model without free adjustable parameters, which will 
 successfully describe the experimental data with  emphasis on the violation of the Koba-Nielsen-Olesen~(KNO) scaling~\cite{KNO}. 

In the first section we start with review of Kharzeev-Levin~(KL) model~\cite{Kharzeev:2017qzs} in zero transverse dimension defining the von Neumann entropy applied to the Pomeron evolution and the moments $C_q$ that can be directly compared to the  experimental data.

In the next section we review the Pomeron evolution equation based on the AGK cutting rules developed  by  Levin and one of the authors~\cite{Levin:2008cgz}~(see also \cite{Levin:2007yv, Kozlov:2007xc, Kormilitzin:2008rk}) and discuss negative probabilities which 
arise in this context. The negative probabilities originate from the negative AGK weights of the relative contributions to the total cross section   and related to the fact that the  AGK cutting rules are derived in the Minkowski space.  We reformulate the Pomeron evolution in the Euclidean space based on the combinatorial considerations used in the original AGK cutting rules. The resulting 
evolution equation correctly reflects the positive definite probabilities of having cut and uncut Pomerons as a function of rapidity. Using the AGK based positive definite probabilities we
construct the von Neumann entropy and calculate $C_q$ as function of rapidity and the mean multiplicity of the produced particles.

 Our model based on the AGK cutting rules successfully describes the  $C_q$ as a function of the mean multiplicity of the produced particles in the experimental data of  the  $\mathtt{p-p}$ collision by  $\mathtt{ALICE}$ Collaboration~\cite{ALICE:2015olq} without any adjustable parameter.
 
 In the last section we summarize and discuss our results. 

\section{Toy model at zero transverse dimension}

 In this section we follow the lines of the dipole evolution analysis for the entanglement in QCD done by  Kharzeev and Levin~\cite{Kharzeev:2017qzs}.
 In this paper we use  Pomeron as  the proper degree of freedom and adopt the notation of the paper by Levin and one of the authors~\cite{Levin:2008cgz}.
It is convenient to define the generating function
\begin{eqnarray}\label{Z0def}
Z_0(u|Y)=\sum_{n=1}^\infty P_n(Y) u^n.
\end{eqnarray}
that introduces $P_n(Y)$,  the probability  of having $n$ Pomerons at a given rapidity $Y$.
The definition of the generating function implies that 
\begin{eqnarray}\label{Z0unity}
Z_0(u=1|Y)=1
\end{eqnarray}
if the total probability is properly normalized. 

The Markov chain equation for the Pomeron evolution reads
\begin{eqnarray}
 \frac{\partial P_n(Y)}{\partial Y}= -\alpha n P_n(Y) + \alpha (n-1) P_{n-1}(Y)
 \label{markov}
\end{eqnarray}

The equation in \eqref{markov} can be written as 
\begin{eqnarray}
 \frac{\partial Z_0(u|Y)}{\partial Y} = -\alpha \; u (1-u) \frac{\partial Z_0(u|Y)}{\partial u} 
 \label{Z0eq}
\end{eqnarray}
where $\alpha$ denotes the Pomeron splitting vertex. 
This evolution  equation is solved for the initial condition 
\begin{eqnarray}\label{initZ0}
Z_0(u|Y=0)=u
\end{eqnarray}
that corresponds to starting with one Pomeron at $Y=0$. 
 The solution of \eqref{Z0eq} given by 
\begin{eqnarray}
Z_0(u|Y)=\frac{u }{u +(1-u) e^{\alpha Y} }
\label{Z0usol}
\end{eqnarray}
satisfies the initial condition in \eqref{initZ0} and the boundary condition in \eqref{Z0unity}.
Using the definition of $Z_0(u|Y)$ in  \eqref{Z0def} one can extract the probability $P_n(Y)$ of having $n$ Pomerons at rapidity $Y$
\begin{eqnarray}
P_n(Y)= e^{-\alpha Y} (1- e^{-\alpha Y})^{n-1}
\label{Pn}
\end{eqnarray}
It is worth emphasizing that $P_n(Y)$ in \eqref{Pn} represents the probability of the \textit{Geometric Distribution} also known as the  \textit{Furry Distribution}~\cite{distribution}
\begin{eqnarray}\label{Pgeom}
\mathtt{Pr} \mathtt{ (X=k)}=p \;(1-p)^{k-1} , \;\;\; k=1,2,3,... 
\end{eqnarray} 
 $\mathtt{Pr}  \mathtt{ (X=k)}$ stands for  the probability of having   the first occurrence of success after $k$ independent trials in the  Bernoulli process, each with success probability $p$.
The Geometric Distribution is the only discrete probability distribution which is \textit{memoryless}, i.e. previous fails does not affect future trials required for success.  Another interesting property of the Geometric Distribution relevant for our discussion is that out of all possible known discrete probability distributions it gives the maximal entropy for the same set of parameters. 
 The  Bernoulli process is the process of having equally distributed binary variable ( $0$- "fail" and $1$-"success"). In our case, the meaning of "fail" is the absence of Pomeron at  rapidity $Y$, where it potentially could be if we had only splitting at each stage of evolution (the $u^2 \frac{\partial }{\partial u}$ term  in \eqref{Z0eq}). The      $u \frac{\partial }{\partial u}$ term in \eqref{Z0eq} is responsible for  propagation of a Pomeron without splitting thus introducing a "vacancy" or the  absence of a Pomeron, which we recognize as "fail" with probability $q=1-p$, where $p=e^{-\alpha Y}$.

The mean number of Pomerons at rapidity $Y$ is given by
\begin{eqnarray}
\langle n \rangle=\sum_{n=1}^\infty  n P_n(Y)= e^{\alpha Y}
\end{eqnarray}
 which can be used to express the probability in \eqref{Pn} in terms of $\langle n \rangle$
\begin{eqnarray}\label{Pnn}
P_n\left(Y=\frac{\ln \langle n \rangle}{\alpha}\right)= (\langle n \rangle-1)^{n-1}\langle n \rangle.
\end{eqnarray}

The useful measure of disorder or uncertainty of the system is the entropy. There are a lot of expressions for entropy derived based on its definition in the classical thermodynamics. We adopt the von Neumann entropy
\begin{eqnarray}\label{vonentropy}
S=- \sum_{n=1}^\infty P_n(Y) \ln P_n(Y)
\end{eqnarray}
which measures  the statistical uncertainty of  quantum systems. Plugging the probability in \eqref{Pn} we obtain
\begin{eqnarray}
S=  \alpha Y e^{\alpha Y} -(e^{\alpha Y}-1) \log (e^{\alpha Y}-1),
\end{eqnarray}
which is a large rapidity can be approximated by 
\begin{eqnarray}
S \simeq \alpha  Y+1-\frac{e^{-\alpha  Y}}{2}=\ln  \langle n \rangle +1 -\frac{1}{2 \langle n \rangle}.
\end{eqnarray}
The fact that the entropy equals the log of the mean number of Pomeron states indicates that at $Y \to \infty$
the system is in the most entangled state. 
It is convenient to introduce a universal measure of uncertainty that can be used for comparison of various systems with different degrees of freedom or alternative states. 

Inspired by  the \textit{efficiency index} for  communication channels  in the work of 
  Wilcox~\cite{Wilcox}
we introduce the \textit{uncertainty  index}
\begin{eqnarray}\label{eta}
\eta(Y) =\frac{S}{  \ln \langle n \rangle}=\frac{-\sum_{n=1}^\infty P_n(Y) \ln P_n(Y)}{ \ln \langle n \rangle }
\label{etadef}
\end{eqnarray}
which measures the  uncertainty (entropy) of the system  normalized by $\ln \langle n \rangle$ where   $\langle n \rangle$  is the average number of Pomerons at a given rapidity $Y$. In other words,  one defines
 $\eta(Y)$  as a relative uncertainty or \textit{the uncertainty indicator}\footnote{The original name \textit{the uncertainty index} of  $\eta(Y)$ relates to the efficiency of data transfer in the information science and could be misleading in our case. Note also the difference in definition  of $\eta(Y)$ where we use $\langle n \rangle$ instead of total number of alternatives in the information science.} that indicates to which extent  different alternatives in the system are equally likely.  In the limiting case of the absolute certainty, i.e. all alternative states are equally likely, the  uncertainty index $\eta(Y)$ approaches unity from above. For $\eta(Y)=1$ all alternative states has the same probability, which makes the system less diverse and thus more certain with maximal entanglement.  The uncertainty   index   $\eta(Y)$  in 
\eqref{eta} 
 can be viewed as  a reasonable entanglement measure for relative comparison of quite different systems. 
 
For the probability given in \eqref{Pn} it reads
\begin{eqnarray}\label{etau}
\eta(Y)= e^{\alpha  Y}-\frac{\left(e^{\alpha  Y}-1\right) \ln \left(e^{\alpha  Y}-1\right)}{\alpha  Y}=1-\frac{1-e^{-\alpha Y }}{ e^{-\alpha Y}}\frac{  \ln  \left(1-e^{-\alpha Y}\right)}{\alpha Y}
\end{eqnarray}
The  plot of $\eta(Y)$   depicted in  Figure~\ref{fig:eta1} demonstrates convergence of the uncertainty  index to unity at large rapidity
\begin{equation}
\eta(Y) \simeq 1 +\frac{1}{\alpha  Y}  -\frac{e^{-\alpha  Y }}{2 \alpha  Y}= 1+\frac{1}{\ln \langle n \rangle }- \frac{1}{2 \langle n \rangle \ln \langle n \rangle}, \;\;\; Y \to \infty
\end{equation}

\begin{figure}[ht]
\centering
\includegraphics[scale=0.5]{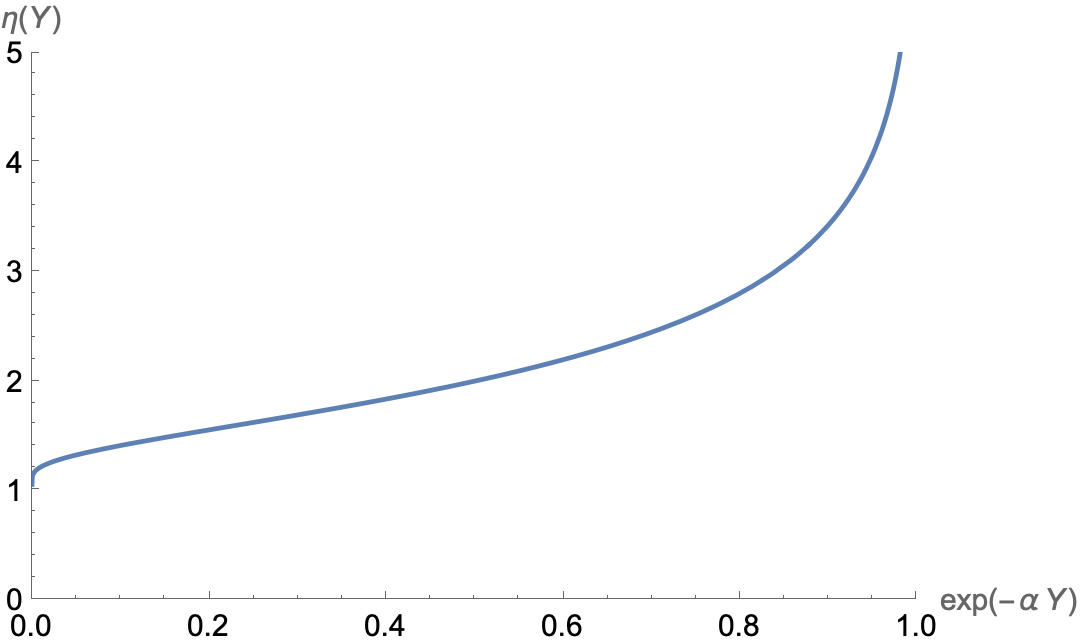}
\caption{ The uncertainty index  $\eta(Y)$ derived from the probability  \newline $P_n(Y)= e^{-\alpha Y} (1- e^{-\alpha Y})^{n-1}$ in \eqref{Pn}  as a function of rapidity $Y$. The uncertainty index  $\eta(Y)$ approaches unity at large rapidity indicating the maximally entangled state. }
\end{figure}\label{fig:eta1}
The uncertainty index  $\eta(Y)$ is a very useful indicator for QCD calculations that makes it possible to use the same expression for the probability of alternative states starting at some rapidity. That would  significantly simplify the involved QCD calculations.

In the next section we extend this analysis to the case of the process of different multiplicities of the produced particles using the Abramovskii-Gribov-Kancheli~(AGK) cutting rules~\cite{AGK}.

\newpage

\section{Entropy and AGK cutting rules}

\begin{figure}[h]
\centering
 \begin{subfigure}{.5\textwidth}
\includegraphics[height=5cm]{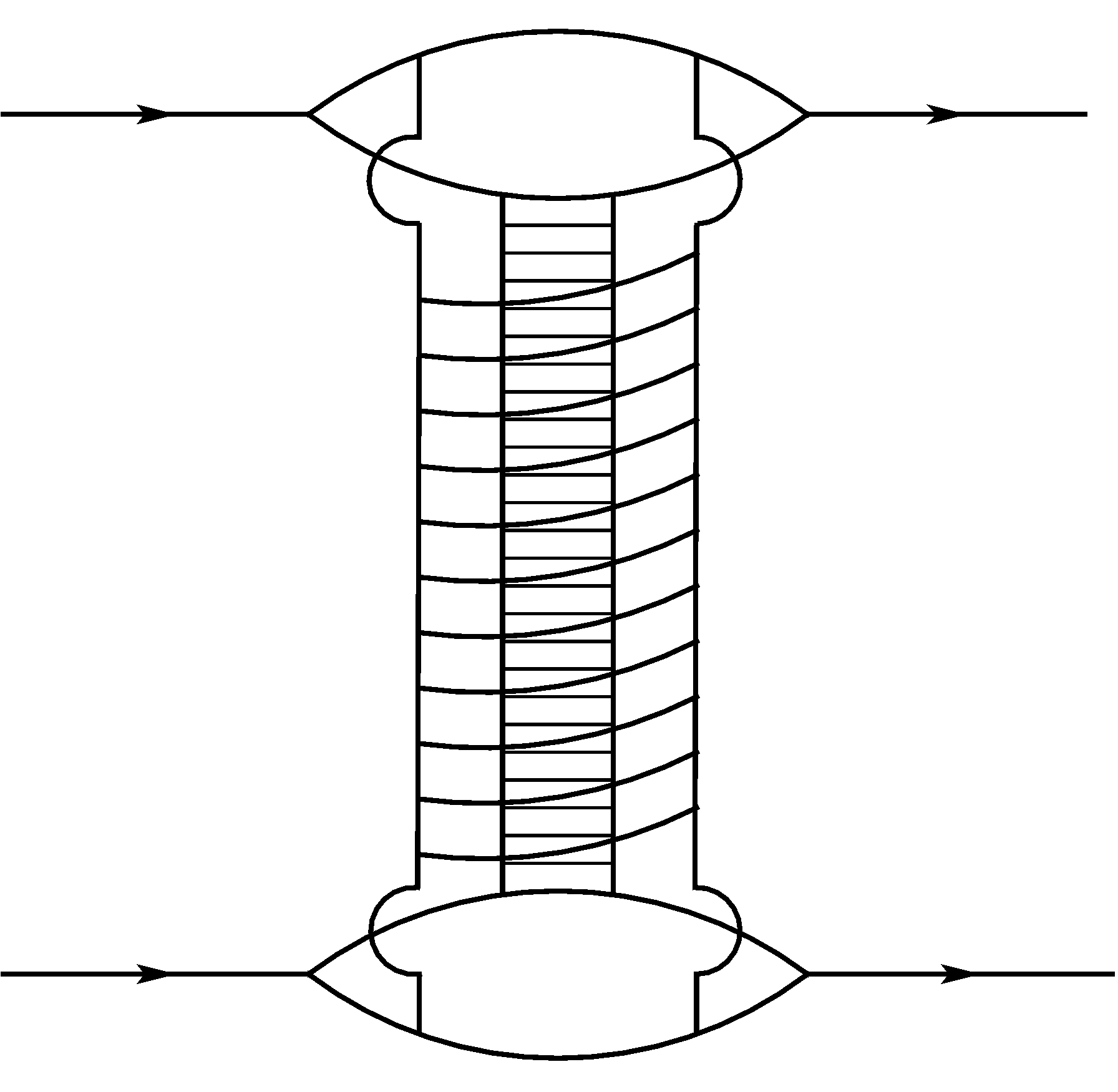}
\caption{Mandelstam cross as cylinder }
\end{subfigure}%
 \begin{subfigure}{.5\textwidth}
\includegraphics[height=5cm]{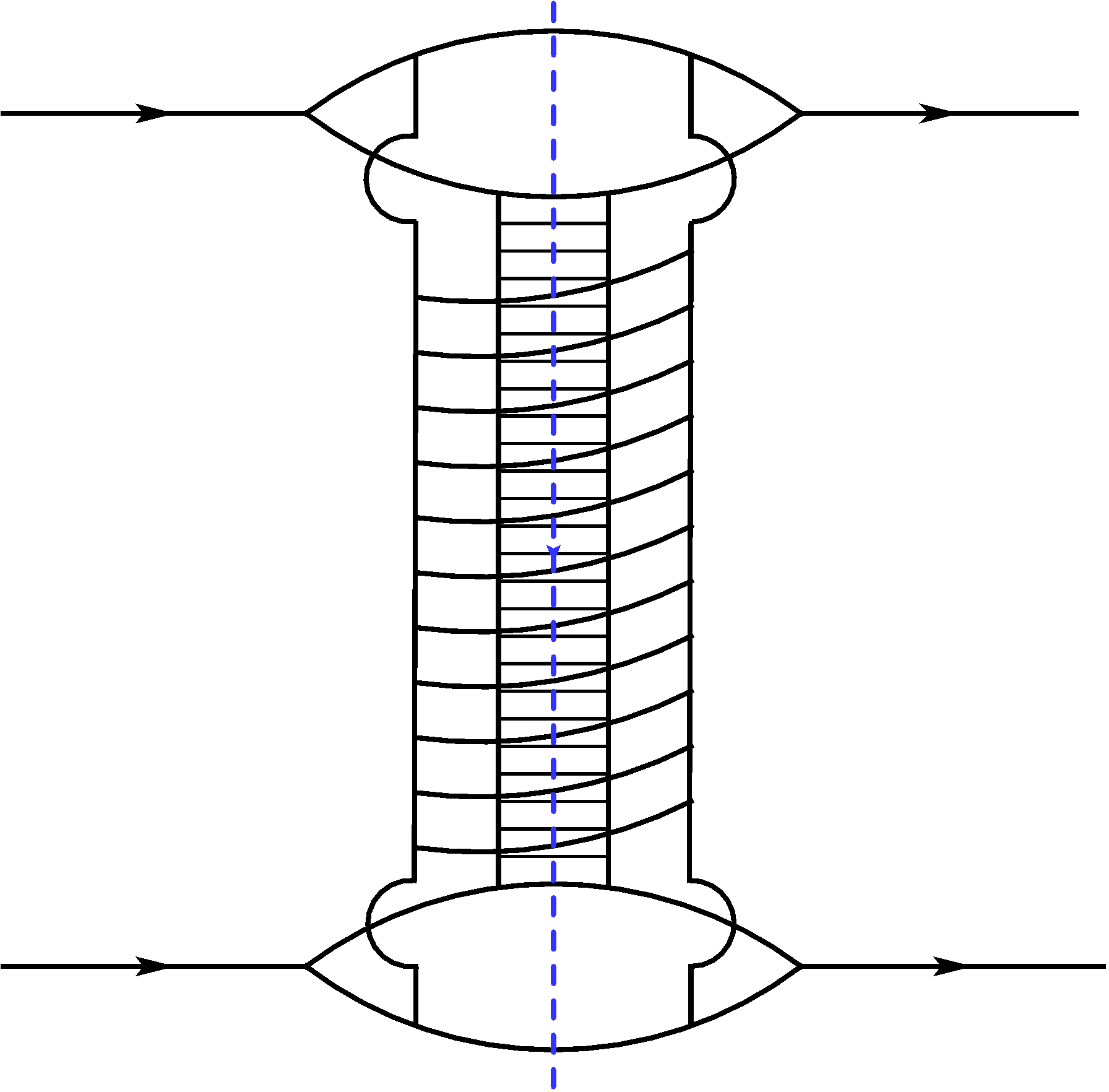}
\caption{Mandelstam cross with double cut }
\end{subfigure}
\caption{The cylindrical topology of   two Pomeron interaction with Mandelstam cross. Two Pomerons can be simultaneously cut putting the produced particles on mass shell.    }
\label{fig:cylinder}
 \end{figure}

Abramovskii-Gribov-Kancheli~(AGK) cutting rules~\cite{AGK} were originally derived to explain the relative contributions of different multiplicity of produced particles to the total cross section. The AGK paper deals with two Pomeron exchange, where one, two or none of the Pomerons can be "cut", i.e. the produced particles that build the Pomeron ladder diagrams are put "on-shell" leading to real production.
The AGK cutting rules  can be summarized as relative weights of the diffraction term~(two uncut Pomerons), single multiplicity term~(one cut and one uncut Pomeron) and the double multiplicity term~(two cut Pomerons).
Those weight and their signs are result of combinatorics and loop integration of the corresponding diagrams. 
It is important to review the origin of those weights in order to understand how to apply them to evolution equation. 
\begin{figure}[ht]
\centering
 \begin{subfigure}{.5\textwidth}
\includegraphics[height=4.5cm]{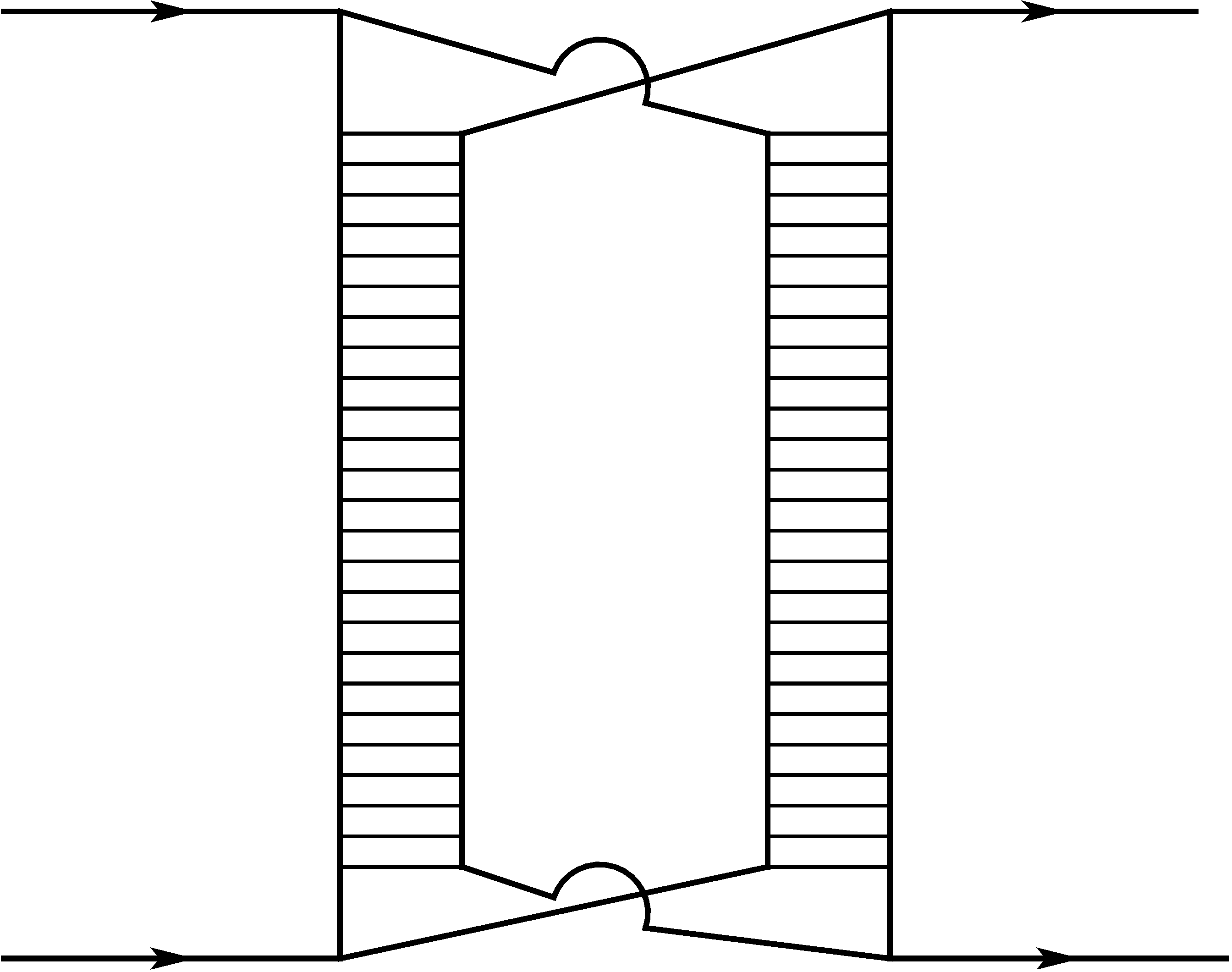}
\caption{Mandelstam cross }
\end{subfigure}%
 \begin{subfigure}{.5\textwidth}
\includegraphics[height=4.5cm]{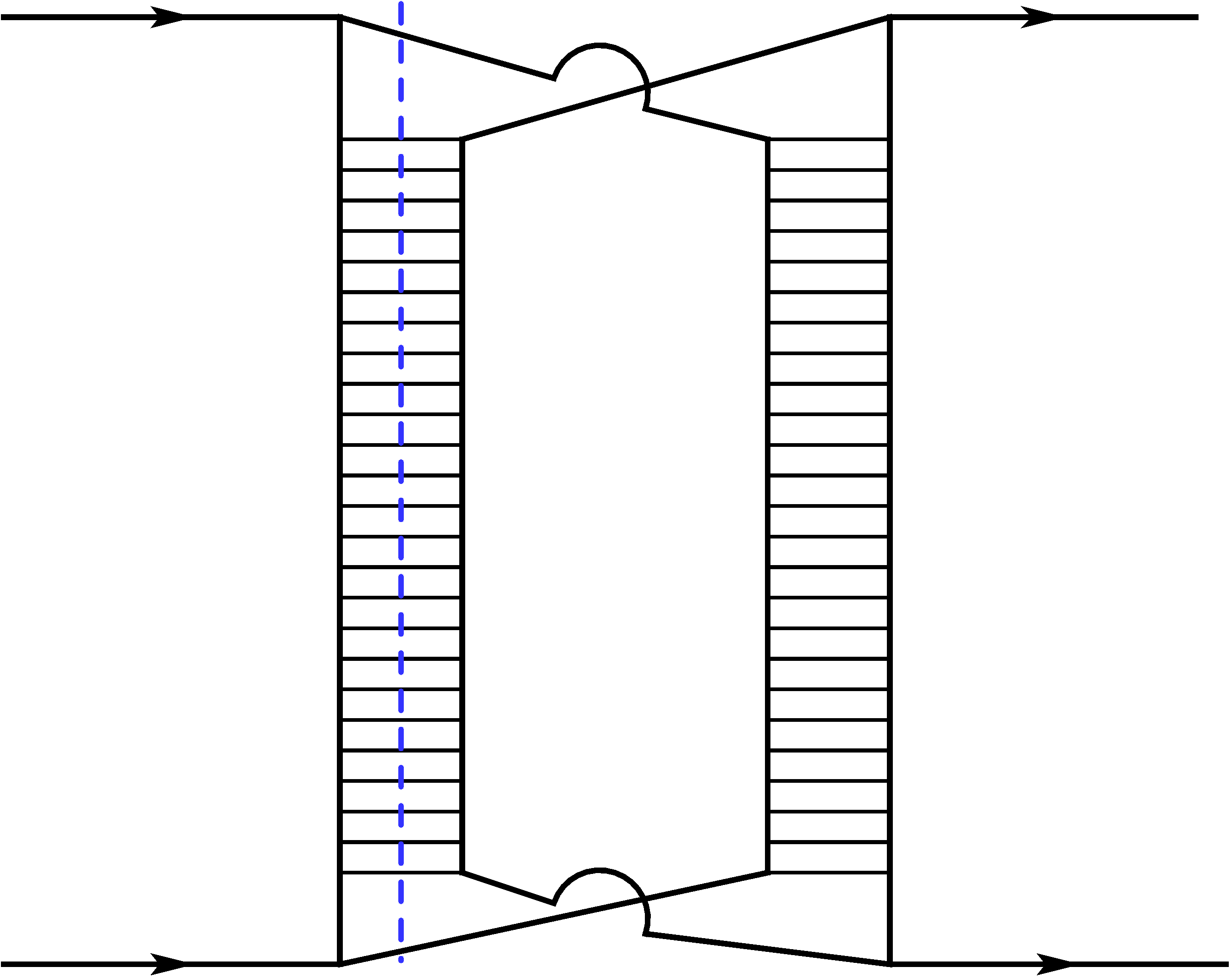}
\caption{Mandelstam cross with one cut }
\end{subfigure}
\caption{Two Pomeron interaction with Mandelstam cross, which allows two Pomerons to be cut simultaneously due to the  non-planar cylindrical  topology.   }
\label{fig:mandel}
 \end{figure}
Fig.~\ref{fig:mandel} shows  a coupling of two Pomerons to the
target and the projectile, which is known as the Mandelstam cross. The Mandelstam cross coupling  allows to put simultaneously on mass shell the intermediate produced particles in both Pomerons due to the cylindrical topology depicted in 
Fig.~\ref{fig:cylinder}. 
  
The AGK cutting rules determine the relative weights of the contribution of different multiplicity of the produced particles to the total cross section in two Pomeron exchange. 
Those weights are derived from combinatorics of all possible cuts and unitarity considerations, where the discontinuity of the amplitude of  each cut Pomeron equals two times the 
imaginary part of scattering amplitude of the one Pomeron exchange 
\begin{eqnarray}
i\delta A=i 2\; \mathtt{Im}A
\end{eqnarray}
The combinatorial coefficients, on the other hand, treat the cut and uncut Pomerons on  equal basis  as follows. For the contribution of diffractive cut~(no cut Pomerons) the combinatorial coefficient is $2$ due to two possibilities of having one or another Pomeron in the amplitude or complex conjugate amplitude as shown in Fig.~\ref{fig:mandel}.

The AGK cutting rules used by one of the authors~\cite{Levin:2008cgz} to extend the analysis of the dipole model in the generating function approach for zero transverse dimension~\cite{Levin:2007yv, Kozlov:2007xc,Kormilitzin:2008rk}. 
The resulting differential equation reads 
\begin{eqnarray}
\frac{\partial Z(w,\bar{w},v|Y)}{\partial Y}&=& -\alpha \left\{ w (1-w)\frac{\partial Z(w,\bar{w},v|Y)}{\partial w}+\bar{w} (1-\bar{w})\frac{\partial Z(w,\bar{w},v|Y)}{\partial \bar{w}}\right\}
\nonumber
\\
&& -\alpha \left\{ 2 w \bar{w}-2 w v - 2\bar{w} v  +v^2 +v\right\} \frac{\partial Z(w,\bar{w},v|Y)}{\partial v}
\label{Zwbarwveq}
\end{eqnarray}
where $\alpha=\Gamma(1 \to 2)$ is the triple Pomeron vertex and the generating function is defined by 
\begin{eqnarray}
Z(w,\bar{w},v|Y)=\sum_{i=0}^\infty  \sum_{j=0}^\infty  \sum_{k=0}^\infty P_{i,j,k}(Y) w^i \bar{w}^j v^k
\label{Zwbarwvdef}
\end{eqnarray}
The equation in \eqref{Zwbarwveq} can diagonalized in variables $w$, $\bar{w}$ and $\xi= w+\bar{w}-v$ resulting in 
\begin{eqnarray}
&&\frac{\partial Z(w,\bar{w},\xi-w-\bar{w}|Y)}{\partial Y}=
\\
&& -\alpha \left \{ \; w (1-w)\frac{\partial   }{\partial w} +\bar{w} (1-\bar{w})\frac{\partial   }{\partial \bar{w}} 
\right \} Z(w,\bar{w},\xi-w-\bar{w}|Y)
\nonumber 
\\
&&  -\alpha \; \xi (1-\xi)\frac{\partial }{\partial \xi}   Z(w,\bar{w},\xi-w-\bar{w}|Y)
\label{Zwbarwxi}
\end{eqnarray}
The solution is a linear combination of three generating functions 
\begin{eqnarray}
Z(w,\bar{w},v|Y)= \mathtt{C}_1 \; Z_0(w|Y) +\mathtt{C}_2 \; Z_0(\bar{w}|Y)+\mathtt{C}_3 \; Z_0(w+\bar{w}-v|Y)
\label{Zwbarwxi}
\end{eqnarray}
where $Z_0(u|Y)$ is the solution to the simple toy model given in \eqref{Z0usol}.
The initial condition 
\begin{eqnarray}
Z(w,\bar{w}, v|Y=0)=v,
\end{eqnarray}
which  corresponds to having one cut Pomeron at $Y=0$ implies that  the solution  in \eqref{Zwbarwxi} is given by 
\begin{eqnarray}
Z(w,\bar{w},v|Y)=   Z_0(w|Y) +  Z_0(\bar{w}|Y)-  Z_0(w+\bar{w}-v|Y).
\label{ZwbarwxiInit}
\end{eqnarray}
The generating function in \eqref{ZwbarwxiInit} solves the differential equation in \eqref{Zwbarwveq} and can be used for deriving scattering amplitude for various processes. 
However, it follows directly from the definition of the generating function and the relative signs of different AGK terms in   the differential equation in \eqref{Zwbarwveq} that in some cases the expression in \eqref{ZwbarwxiInit} leads to negative probabilities, which are perfectly allowed in Quantum Field Theory, where the AGK cutting rules were derived. 
Any attempt of using the QFT results directly in Quantum Mechanics would fail due the presence of the negative probabilities that have no proper interpretation in Quantum Mechanics.  
One of such applications is the use of the concept of  von Neumann entropy which requires, by definition, all probabilities to be positive definite. The von Neumann entropy is defined by 
\begin{eqnarray}
S=-\mathtt{Tr} (\rho \ln \rho),
\label{entropyrho}
\end{eqnarray}
 where the probability density matrix 
 \begin{eqnarray}
 \rho =\sum_j p_j \ket{j} \bra{j}.
 \end{eqnarray}
Plugging $\rho$ into the definition in \eqref{entropyrho} one obtains 
\begin{eqnarray}
S=-\sum_j p_j \ln p_j.
\end{eqnarray}
 This definition of the entropy implies the well defined space of states $\bra{j}$ and the positiveness of $p_j$. None of those conditions is fulfilled in QFT.  
 
 Nevertheless, it is still possible to apply AGK cutting rules in the modified version for the entanglement analysis using von Neumann entropy. The first step is to understand the origin of the coefficients of different multiplicity terms, $\sigma_n$ that lead to the AGK coefficients 
\begin{eqnarray}
\sigma_0 \div \sigma_1 \div \sigma_2= 2 \div (-8) \div 4
\label{AGKcoeff}
\end{eqnarray} 
 that sum to the total cross section 
 \begin{eqnarray}
 \sigma_0+\sigma_1+\sigma_2= -2 \mathtt{Im} A_1 \mathtt{Im} A_2
 \label{sigmasum}
 \end{eqnarray}
  Those relative weight coefficients are obtained from $a)$ positive definite combinatorial weights of different ways to cut two Pomerons 
 $b)$ a minus sign that appears every time the Pomeron is cut. It is worth emphasizing the latter. The leading contribution of the cut Pomeron is given by 
\begin{eqnarray}
i\delta A=i 2\; \mathtt{Im}A,
\end{eqnarray}
that can be derived by the loop integration with the produces particles put on-shell, in contrast to the uncut Pomeron where the integration is performed calculating the residue at one of the poles of the loop integration over longitudinal momentum. This is the case in the Minkowski space, where the longitudinal component of the loop momentum comes with a minus sign with respect to the transverse components. On contrary, in the Euclidean space all momenta components come at the equal basis and thus the relative sign between the cut Pomeron and the uncut Pomeron contributions does   not appear resulting in the AGK weights in the Euclidean space
\begin{eqnarray}
\sigma^{\mathtt{Euclidean}}_0 \div \sigma^{\mathtt{Euclidean}}_1 \div \sigma^{\mathtt{Euclidean}}_2= 2 \div 8 \div 4
\label{AGKcoeffEucl}
\end{eqnarray}
The immediate question that emerges in this context if the positive definite AGK weights in the Euclidean space can sum into a meaningful expression that can be associated with the total cross section of the two Pomeron exchange, similar to the Minkowski space where the expression in \eqref{sigmasum} holds. In the Euclidean space the sums of all contributions of the different multiplicity in \eqref{AGKcoeffEucl} gives 
 \begin{eqnarray}
 \sigma^{\mathtt{Euclidean}}_0+\sigma^{\mathtt{Euclidean}}_1+\sigma^{\mathtt{Euclidean}}_2= 10 \mathtt{Im} A_1 \mathtt{Im} A_2
 \label{sigmasumEucl}
 \end{eqnarray}
and there is no analytic continuation from the Euclidean to the Minkowski space that would give the physical total cross section in \eqref{sigmasum}. 
This question has been  well studied in the context of the helicity amplitudes~\cite{Bartels:2008ce,Bartels:2009vkz,Lipatov:2010qg, Lipatov:2010ad,Bartels:2010tx,Bartels:2011ge} and it was shown that the operation of the analytic continuation  from the Euclidean to the Minkowski space does not commute with the operation of taking Regge limit, i.e. considering the  multi-Regge kinematics.  This non interchangeability of analityc continuation and the Regge limit is not relevant  in the context of the entropy and the entanglement
because the  statistics of  different multiplicities contributions is correctly represented by counting the number of cut Pomerons and uncut Pomerons with positive definite AGK weights in the Euclidean space in  \eqref{AGKcoeffEucl}. 

In other words, it is perfectly possible to use the modified AGK weights in the Euclidean space in \eqref{AGKcoeffEucl} to count number of cut and uncut Pomerons, but it is not  possible to 
directly apply them to calculation of the total cross section due to non-interchangeability of   the operation of taking Regge limit  
 the operation of the analytic continuation from the Euclidean space to the Minkowski space.
The differential evolution equation based on the AGK weights in \eqref{Zwbarwveq} should be modified in the Euclidean space as follows
\begin{eqnarray}
&&\frac{\partial Z^{Eucl}(w,\bar{w},v|Y)}{\partial Y}= \nonumber
\\
&&
 -\alpha \left\{ w (1-w)\frac{\partial Z^{Eucl}(w,\bar{w},v|Y)}{\partial w}+\bar{w} (1-\bar{w})\frac{\partial Z^{Eucl}(w,\bar{w},v|Y)}{\partial \bar{w}}\right\}
\nonumber
\\
&& +\alpha \left\{ 2 w \bar{w}+2 w v + 2\bar{w} v  +v^2 -v\right\} \frac{\partial Z^{Eucl}(w,\bar{w},v|Y)}{\partial v}
\label{Zwbarwveq}
\end{eqnarray} 
which is solved by 
\begin{eqnarray}
Z^{Eucl}(w,\bar{w},v|Y)= \mathtt{C}_1 \; Z_0(w|Y) +\mathtt{C}_2 \; Z_0(\bar{w}|Y)+\mathtt{C}_3 \; Z_0(w+\bar{w}+v|Y)
\label{Zwbarwxi}
\end{eqnarray}
where $Z_0(u|Y)$ is the solution  to the simple toy model (see \eqref{Z0usol})
\begin{eqnarray}
Z_0(u|Y)= \frac{u e^{-\alpha Y}}{1+u(e^{-\alpha Y}-1)}.
\label{Z0usol2}
\end{eqnarray}

In this paper we discuss a process where there is at least one cut Pomeron. In this case, the only relevant terms in \eqref{Zwbarwxi} is the term that depends on $v$, namely, $Z_0(v+w+\bar{w}|Y)$.
Using the definition of the generating function 
\begin{eqnarray}
Z_0(v+w+\bar{w}|Y)&=&\sum_{i,j,k=0}^\infty P_{i,j,k}(Y) v^i w^j \bar{w}^k 3^{i+j+k} (1- \delta_{i,0}\delta_{j,0} \delta_{k,0}) 
\\
&&=\frac{v+w+\bar{w}}{ v+w+\bar{w}+(1- v-w-\bar{w})e^{\alpha Y}} \nonumber
\label{Z03}
\end{eqnarray}
we extract the probability
\begin{eqnarray}\label{pijk}
P_{i,j,k}(Y)= \frac{1}{3^{i+j+k}}\frac{(i+j+k)!} {i! j!  k! }  e^{-\alpha Y} (1 - e^{-\alpha Y})^{i+j+k-1}
\label{Pijk}
\end{eqnarray}
Note the  $3^{i+j+k}$ term in \eqref{Z03} introduced as a proper normalization due to the fact that the argument $w+\bar{w}+v$ equals $3$  at $w=\bar{w}=v=1$ in contrast to the definition of $Z_0(u|Y)$ in \eqref{Z0def}. The first two terms in \eqref{Zwbarwxi} are not relevant for extracting the probability for having at least one cut Pomeron as they do not make any contribution to the cut Pomerons.

The probability $P_{i,j,k}$ is symmetrical with respect to all its indices and its represent a probability of finding  $i$ cut Pomerons, $j$ uncut Pomerons in the amplitude and $k$ uncut Pomerons in the complex conjugate amplitude at any given rapidity $Y$.

We are interested in a case, where one has at least one cut Pomeron at any given rapidity $Y$ limiting the total number of Pomerons to $m=i+j+k$.

 We define  the probability $P_{m}$ of having $m$ Pomerons of any kind under condition that at least one of them is the cut Pomeron, i.e.
\begin{eqnarray}
P_{m}(Y)=\sum_{i=1}^{m} \sum_{j=0}^{m-i} \sum_{k=0}^{m-i-j} P_{i,j,k} (Y)
\end{eqnarray}
it reads
\begin{eqnarray}\label{pmfirst}
P_{m}(Y)&=&  e^{-\alpha Y} (1 - e^{-\alpha Y})^{m-1} \frac{m!}{3^m}  \sum_{i=1}^{m} \sum_{j=0}^{m-i}\frac{1} {i! j!  (m-i-j)!  }  
\\
&=&
  e^{-\alpha Y} (1 - e^{-\alpha Y})^{m-1}\frac{m!}{3^m}  \sum_{i=1}^{m}   \frac{  2^{m-i}}{i! (m-i)! 3^m } \nonumber
\end{eqnarray}
leading to 
\begin{eqnarray}
P_{m}(Y)= \left(1-\frac{2^m}{3^m} \right) e^{-\alpha Y} (1 - e^{-\alpha Y})^{m-1}   \nonumber
\end{eqnarray}
Note that $P_{m}(Y)$   does not account for cases  of having only $m$ uncut Pomerons and thus the sum $\sum_{m=1}^\infty P_m(Y)$ is not strictly equal to unity, i.e.
\begin{eqnarray}
 \sum_{m=1}^\infty P_m(Y) = \frac{1}{1+2 e^{-\alpha Y}}
\end{eqnarray}
which goes to unity when $Y\to \infty$ because  the contribution of those states at large rapidity  is negligible. 
According to the boundary condition of having at least one cut Pomeron at any rapidity we exclude those states normalizing $P_{m}(Y)$ as follows
\begin{eqnarray}
\tilde{P}_m(Y) =\frac{P_m (Y)}{ \sum_{m=1}^\infty P_m(Y) }
\end{eqnarray}
resulting in 
\begin{eqnarray}
\tilde{P}_m(Y) =\left(1-\frac{2^m}{3^m} \right) e^{-\alpha Y} (1 - e^{-\alpha Y})^{m-1}(1+2 e^{-\alpha Y})
\label{Pmtilde}
\end{eqnarray}
The normalized probability $\tilde{P}_m(Y) $ sums to unity 
\begin{eqnarray}
\sum_{m=1}^\infty \tilde{P}_m(Y) =1
\end{eqnarray}
and satisfies the initial condition 
\begin{eqnarray}
\tilde{P}_m(Y=0)=\delta_{m,1}.
\end{eqnarray}

This should be compared to the probability used by Kharzeev and Levin~(the KL model)~\cite{Kharzeev:2017qzs}
 \begin{eqnarray}
P^{KL}_m(Y) =  e^{-\alpha Y} (1 - e^{-\alpha Y})^{m-1}
\label{PmKL}
\end{eqnarray}
used for calculating the entropy and moments $C_q$ for comparison to the experimental data.  
We follow the lines of the  KL paper~\cite{Kharzeev:2017qzs} as well as the more recent works~\cite{Kharzeev:2021yyf,Zhang:2021hra,Florio:2023dke,Hentschinski:2023izh,Hentschinski:2023nhy,Kutak:2023cwg,Kutak:2024pfy,Hatta:2024lbw,Levin:2024wtl,Kutak:2025hzo,Moriggi:2024tbr,Caputa:2024xkp,Bhattacharya:2024sno,Liu:2022bru}
and calculate the experimental observables using the probability $\tilde{P}_m(Y)$ in \eqref{Pmtilde} based on the   AGK cutting rules.  
The mean multiplicity of produced particles corresponds  the mean number of Pomerons that is given by 
\begin{eqnarray}
<m>=\sum_{m=1}^\infty m \tilde{P}_m(Y)=e^{\alpha  Y}-1+\frac{3}{1+2 e^{-\alpha  Y}},
\end{eqnarray}
and  at large rapidity takes form 
\begin{eqnarray}
<m>\simeq e^{\alpha  Y}+2-6 e^{-\alpha  Y}.
\end{eqnarray}

The entropy is defined as  
\begin{eqnarray}\label{entropy}
S=-\sum_{m=1}^\infty \tilde{P}_m(Y) \ln \tilde{P}_m(Y)
\end{eqnarray}
and can be reasonably  approximated and found analytically  using 
\begin{eqnarray}
\log \left(1- \frac{2^m}{3^m}  \right) =- \sum_{k=1}^\infty \left(\frac{2^m}{3^m} \right)^{  k },
\end{eqnarray}
where we limit ourselves to the first hundred terms.  Plugging this in \eqref{entropy} for large rapidity  we obtain the first three leading terms of the entropy at high energies
\begin{eqnarray}
S=-\sum_{m=1}^\infty \tilde{P}_m(Y) \ln \tilde{P}_m(Y) \simeq  \alpha Y +1 - e^{-\alpha Y}
\end{eqnarray}

\begin{figure}[ht]
\centering
\includegraphics[scale=0.5]{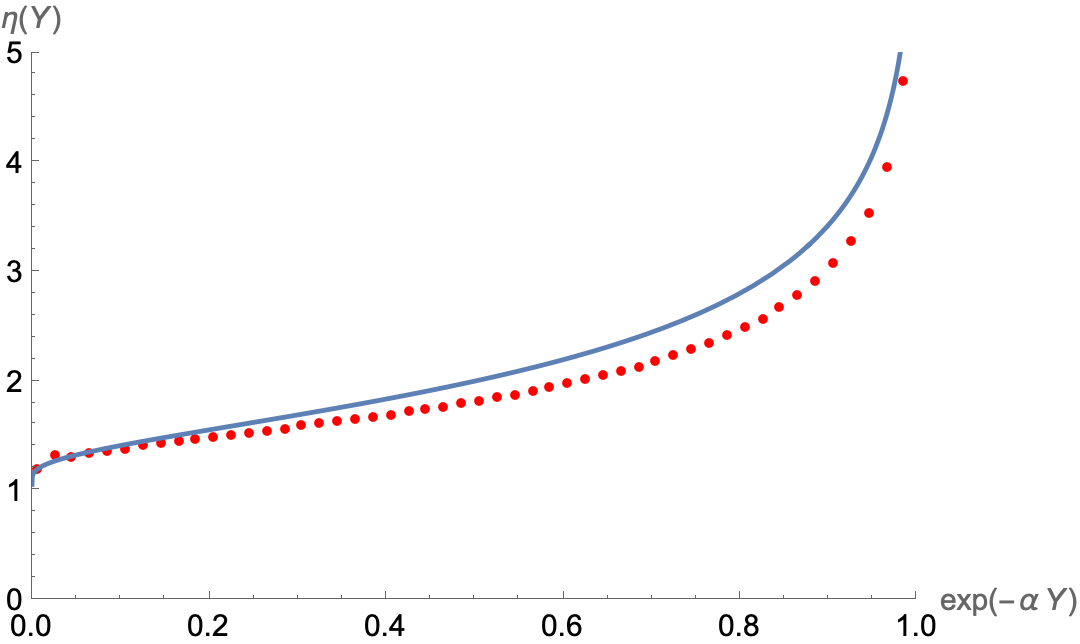}
\caption{ Comparison of the  uncertainty index  $\eta(Y)$ derived from the probability in Kharzeev-Levin model~(solid line) and our model~(the AGK model)~(red dotted line).   The uncertainty index  $\eta(Y)$ approaches unity at large rapidity indicating the maximally entangled state. }
\label{fig:eta12}
\end{figure}

At $Y=0$ the entropy vanishes due to the fact that $\tilde{P}_m(Y=0)= \delta_{m,1}$. Vanishing entropy at $Y=0$ implies that the system is in the pure state and thus  it is maximally ordered. 
On the other hand,  at  large rapidity the entropy grows as the mean number of Pomerons  
\begin{eqnarray}
S \simeq \alpha Y \simeq \ln \langle m \rangle
\end{eqnarray}
  implying maximally entangled state. 
The entropy of the  Kharzeev-Levin model has the  form 
\begin{eqnarray}
S_{KL}=\alpha  Y-e^{\alpha  Y} \left(1-e^{-\alpha  Y}\right) \log \left(1-e^{-\alpha  Y}\right) \simeq \alpha  Y+1-\frac{e^{-\alpha  Y}}{2}
\end{eqnarray}
leading to the same asymptotics
\begin{eqnarray}
S_{KL} \simeq \alpha Y \simeq \ln \langle m \rangle
\end{eqnarray} 
The uncertainty  index $\eta(Y) $  defined in \eqref{etadef} is also of the  similar behaviour for the two models as one can see from their comparison in Fig.~\ref{fig:eta12}.

 \newpage

\section{Comparison to Experimental Data} 
In the framework of the Koba-Nielsen-Olesen~(KNO)~\cite{KNO}   scaling analysis, it is assumed that at high energies the probability of  having $n$ Pomerons should scale as
\begin{eqnarray}
P_n(Y)=\frac{1}{\langle n \rangle} \Psi \left( \frac{n}{\langle n \rangle}  \right),
\end{eqnarray}
where $\Psi(x)$ is a smooth bounded analytic function and  $\langle n \rangle$ is the average multiplicity of the produced particles. If the KNO scaling holds the $q$-moments $C_q$ are defined by 
  \begin{eqnarray}
  C_q=\frac{\langle n^q \rangle }{\langle n \rangle ^q},
  \label{Cndef} 
  \end{eqnarray}
  should be energy independent, 
 because 
  \begin{eqnarray}
  \langle n^q \rangle =\sum_{n=1}^\infty n^q P_n(y) 
  \end{eqnarray}
  can be approximated by the integral
\begin{eqnarray}
\langle n^q \rangle =\int \frac{n^q}{\langle n \rangle} \Psi \left( \frac{n}{\langle n \rangle}  \right)  d n =\langle n \rangle ^q \;\mathtt{Const_q}
\label{cqconst}
\end{eqnarray}  
  where $\mathtt{Const_q}$ is some number, which does not depend on energy. All energy dependence is scaled in $\langle n \rangle^q$
and in accordance to the definition of the moments $C_q$ it cancels with the denominator provided the KNO scaling holds, namely  plugging \eqref{cqconst} into \eqref{Cndef} we obtain
\begin{eqnarray}
  C_q=\frac{\langle n^q \rangle }{\langle n \rangle ^q}=\frac{\langle n \rangle ^q \;\mathtt{Const_q}}{\langle n \rangle ^q}=\mathtt{Const_q}.
  \label{Cqcancel} 
  \end{eqnarray}  
  This implies that the moments $C_q$ are independent of energy within limits of the KNO scaling conditions.

  In order to check the validity of our model we calculate the moments $C_q$ using the AGK based probability in \eqref{Pmtilde} and compare it to the experimental data. 
 The probability calculated in our model using the AGK cutting rules in \eqref{Pmtilde} is given by  
 \begin{eqnarray}
\tilde{P}_n(x) =\left(1-\frac{2^n}{3^n} \right) x  (1 - x)^{n-1}(1+2 x)
\label{pnAGK}
\end{eqnarray} 
  where $x=e^{-\alpha Y}$. 
  
  The average number of Pomerons for the AGK probability reads 
\begin{eqnarray}
 \left.\langle n^q \rangle \right|_{q=1} =\langle n \rangle =\frac{3}{2 x+1}+\frac{1}{x}-1
 \label{averagen1}
  \end{eqnarray}

In the general case of natural $q$   the numerator in \eqref{Cndef} can be written as follows
\begin{eqnarray}
  \langle n^q \rangle =\frac{x (1+2 x)}{1-x} \left[ \mathtt{Li}_{-q}(1-x)-\mathtt{Li}_{-q}\left(\frac{2 (1-x)}{3}\right)\right],
  \label{na}
\end{eqnarray}   
  where $x=e^{-\alpha Y}$ and $\mathtt{Li}_{b}(z)$ is the special function named polylog function~(polylogarithm), which cannot be expressed in terms of the elementary 
  functions for positive integer values of $b>1$.  For the negative integer index  $b$ it is a rational function given by 
\begin{eqnarray}
\mathtt{Li}_{-q}(z)= \left( z \frac{\partial }{ \partial z}\right)^q \frac{z}{1-z}, \;\; q=1,2,3...
\end{eqnarray}  
  The first few polylogarithms of  the negative integer index read
  \begin{eqnarray}
  &&\mathtt{Li}_{-1}(z)=\frac{z}{(1-z)^2},  \\
   &&\mathtt{Li}_{-2}(z)=\frac{z^2+z}{(1-z)^3},\nonumber
   \\
   &&\mathtt{Li}_{-3}(z)=\frac{z^3+4 z^2+z}{(1-z)^4}, \nonumber
   \\
   &&\mathtt{Li}_{-4}(z)=\frac{z^4+11 z^3+11 z^2+z}{(1-z)^5}, \nonumber
   \\
   &&\mathtt{Li}_{-5}(z)=\frac{z^5+26 z^4+66 z^3+26 z^2+z}{(1-z)^6} \nonumber
   \label{polyrational}
  \end{eqnarray}
  
Plugging \eqref{na} and    \eqref{averagen1} into \eqref{Cndef} we obtain

\begin{eqnarray}
C_q=\frac{\langle n^q \rangle}{\langle n \rangle^q } =\frac{x^{q+1} (1+2 x)^{q+1}}{(1-x)(1+4 x -2 x^2)} \left[ \mathtt{Li}_{-q}(1-x)-\mathtt{Li}_{-q}\left(\frac{2 (1-x)}{3}\right)\right]
\label{CqLi}
\end{eqnarray}
Note that the expression $1+4 x -2 x^2$ never turns zero as follows from the definition $x=e^{-\alpha Y}$. The divergent overall factor $1/(1-x)$ is cancelled by highly convergent difference of the 
polylog functions in the brackets in \eqref{CqLi}.
 The first five moments $C_q$ read  
 \begin{eqnarray}
&& C_2=   \frac{ 4 x^4-26 x^3+18 x^2+11 x+2}{\left(2 x^2-4 x-1\right)^2},  \\
&& C_3=  \frac{4 x^4-56 x^3+37 x^2+18 x+6}{\left(2 x^2-4 x-1\right)^2}, \nonumber \\
&& C_4= \frac{16 x^8-568 x^7+2804 x^6-3222 x^5-410 x^4+619 x^3+614 x^2+204 x+24}{\left(2 x^2-4 x-1\right)^4}, 
\nonumber 
 \\
&& C_5= \frac{16 x^8-1168 x^7+8664 x^6-10732 x^5-59 x^4+690 x^3+1830 x^2+720 x+120}{\left(2 x^2-4 x-1\right)^4}   \nonumber
\label{C25x}
 \end{eqnarray}
  or in terms of the average number of Pomerons $\langle n \rangle$ in \eqref{averagen1}

  \begin{eqnarray}
  && C_2=\frac{24+2 \langle n \rangle^2-5 \sqrt{\langle n \rangle^2+24}}{\langle n \rangle^2}, \\
  && C_3=\frac{145+6 \langle n \rangle^2-30 \sqrt{\langle n \rangle^2+24}}{\langle n \rangle^2}, \nonumber \\
&& C_4=\frac{7080+1022 \langle n \rangle^2 +24 \langle n \rangle^4-5 \left(36 \langle n \rangle^2+289\right) \sqrt{\langle n \rangle^2+24}}{\langle n \rangle^4}, 
 \nonumber \\
&&  C_5=\frac{106561+8070 \langle n \rangle^2 +120 \langle n \rangle ^4
  -150 \left(8 \langle n \rangle^2+145\right) \sqrt{\langle n \rangle^2+24}}{\langle n \rangle^4} \nonumber
  \label{C25n}
  \end{eqnarray}

 The moments $C_q$ in  \eqref{C25n} are expressed in terms of the multiplicity $\langle n \rangle$ directly available in the experimental data without any adjustable parameter. This fact allows  a self-consistent direct comparison to the experimental data. We compare our results  to the $\mathtt{p-p}$ data by $\mathtt{ALICE}$ Collaboration~(see Table $13$ in \cite{ALICE:2015olq}).
 The experimental data is given for three pseudorapidity windows $\Delta \eta=0$, $\Delta \eta=1$ and $\Delta \eta=1.5$. From the definition of the pseudorapidity it follows that for our model formulated in zero transverse dimension the pseudorapidity should go to infinity meaning the detector performs the measurement in the entire solid angle of $4 \pi$.  Thus the appropriate available  set of the experimental data for testing our model is for  $\Delta \eta=1.5$. 
 
In  Table~\ref{tab:Veta15} we compare the moments $C_q$ calculated using our model~(AGK model) and the Kharzeev-Levin model~\cite{Kharzeev:2017qzs}~(KL model) to the experimental data for the  $\mathtt{p-p}$ collisions  by $\mathtt{ALICE}$  Collaboration~(see Table $13$ in \cite{ALICE:2015olq}). 
As we have already mentioned, the comparison to the experimental data is done without any adjustable parameter in both models, which makes it possible to compare two different models in the self-consistent way. It is clearly seen from the direct comparison to the experimental data in Table~\ref{tab:Veta15}  that our model based on AGK cutting rules performs significantly better than the KL model in a broad range of the center-of-mass energy.

The main results of this section are given in the attached $\mathtt{Mathematica}$ file.   
\newpage
\begin{table}[h!]
    \centering
    \begin{tabular}{|c|l|l|l|}
    \hline  
  \multicolumn{4}{|c|}{$\sqrt{s}=0.9 \; \mathtt{TeV}$}     \\
     \hline $\mathtt{observable}$ &   $\mathtt{experiment}$    & $\mathtt{AGK \; model}$       &$\mathtt{KL \; model}$  \\
        \hline
        $\langle n \rangle$	& $11.8\pm 0.4$	                  & $ 11.8\pm 0.4$	            &     $ 11.8\pm0.4$		   \\
        $C_2$			       & $1.7\pm0.1$                   	& $1.71 \pm 0.01$	            & $ 1.92\pm0.03$	   \\
         $C_3$			       & $3.9\pm0.3$                   	& $4.29\pm 0.04$	            & $5.50\pm0.02$	   \\
	    $C_4$		         	& $11\pm1$	                       & $14.2\pm0.2$	                & $21.1\pm 0.1$    \\
	   $C_5$		        	& $36\pm6$	                       & $59.0\pm1.1$             	& $100.7\pm0.6$	    \\
        \hline
    \end{tabular}
    \centering
    \begin{tabular}{|c|l|l|l|}
    \hline  
  \multicolumn{4}{|c|}{$\sqrt{s}=2.76 \; \mathtt{TeV}$}     \\
     \hline $\mathtt{observable}$ &   $\mathtt{experiment}$    & $\mathtt{AGK \; model}$       &$\mathtt{KL \; model}$  \\
        \hline
        $\langle n \rangle$	& $14.2\pm 0.7$	                  & $ 14.2 \pm 0.7$	                     & $14.2\pm 0.7$	   \\
        $C_2$			       & $1.8\pm0.1$                   	& $1.75\pm 0.01$	                     & $ 1.93\pm0.01$	   \\
         $C_3$			       & $4.5\pm0.6$                   	& $4.48 \pm 0.05$	                & $ 5.58\pm0.02$	   \\
	    $C_4$		         	& $14\pm3$	                       & $15.3 \pm 0.3$	                  & $21.5\pm0.1$    \\
	   $C_5$		        	& $51\pm 13$	                       & $65.2\pm1.6 $                 	& $103.8\pm0.8$	    \\
        \hline
    \end{tabular}
    \centering
    \begin{tabular}{|c|l|l|l|}
    \hline  
  \multicolumn{4}{|c|}{$\sqrt{s}=7 \; \mathtt{TeV}$}     \\
     \hline $\mathtt{observable}$ &   $\mathtt{experiment}$    & $\mathtt{AGK \; model}$       &$\mathtt{KL \; model}$  \\
        \hline
        $\langle n \rangle$	& $17.5\pm 0.6$	                  & $17.5\pm 0.6$	 	                     & $17.5\pm 0.6$	 	   \\
        $C_2$			       & $1.9\pm0.1$                   	& $1.78\pm 0.01$	                     & $1.94\pm 0.01$	   \\
         $C_3$			       & $5.0\pm0.4$                   	& $4.69 \pm 0.03$	                & $ 5.66\pm0.01$	   \\
	    $C_4$		         	& $16\pm2$	                       & $16.5 \pm 0.2$	                  & $21.99\pm0.07$    \\
	   $C_5$		        	& $64\pm 11$	                       & $72.1\pm1.1 $                 	& $106.8\pm0.4$	    \\
        \hline
    \end{tabular}
    \centering
    \begin{tabular}{|c|l|l|l|}
    \hline  
  \multicolumn{4}{|c|}{$\sqrt{s}=8 \; \mathtt{TeV}$}     \\
     \hline $\mathtt{observable}$ &   $\mathtt{experiment}$    & $\mathtt{AGK \; model}$       &$\mathtt{KL \; model}$  \\
        \hline
        $\langle n \rangle$	& $17.8\pm 1.1$	                  & $17.8\pm 1.1$	 	                     & $17.8\pm 1.1$	 	   \\
        $C_2$			       & $1.9\pm0.1$                   	& $1.78\pm 0.01$	                     & $ 1.94\pm0.01$	   \\
         $C_3$			       & $5.2\pm0.8$                   	& $4.71 \pm 0.06$	                & $ 5.67\pm0.02$	   \\
	    $C_4$		         	& $18\pm4$	                       & $16.5 \pm 0.3$	                  & $22.02\pm0.12$    \\
	   $C_5$		        	& $70\pm 20$	                       & $72.6\pm2.0  $                 	& $107.0\pm0.5$	    \\
        \hline
    \end{tabular}
    \caption{Comparison of the $q$-moments $C_q$ of our model~(AGK model) and  Kharzeev-Levin model~(KL model) to the experimental data for the $\mathtt{p-p}$ collision by $\mathtt{ALICE}$ Collaboration~(see Table $13$ in 
    \cite{ALICE:2015olq}). Here we consider the pseudorapidity $\eta=1.5$ and the center-of-mass energies  $\sqrt{s}=0.9\; \mathtt{TeV}$,  $\sqrt{s}=2.76\; \mathtt{TeV}$,  $\sqrt{s}=7\; \mathtt{TeV}$ and  $\sqrt{s}=8\; \mathtt{TeV}$. Note that we use as the input  the experimental value of the mean multiplicity $\langle n \rangle$ so that the AGK model and the KL model have no freedom in adjusting any  free  parameter  such as  $\alpha_s$ or $Y_0$. Our model~(AGK model) consistently   better than the KL model in describing the experimental data in the wide range of energies.}
    \label{tab:Veta15}
\end{table} 
\newpage


\section{Conclusion and Discussions}\label{}
 
In this paper we analysed the Pomeron evolution based on the AGK cutting rules. A direct application of the AGK rules  to the evolution equation leads to negative probabilities that follow from negative AGK weights derived in the Minkowski space.  We reformulate the Pomeron evolution in the Euclidean space using the combinatorial coefficients derived in the original AGK paper. The resulting positive definite probabilities in \eqref{Pijk} and \eqref{Pmtilde} for cut and uncut Pomerons are then used in calculating the von Neumann entropy and the $q$-moments $C_q$ as a function of rapidity~\eqref{C25x} and a function of the mean multiplicity of the produced particles in \eqref{C25n}. The latter is shown to successfully describe the experimental data for $\mathtt{p-p}$ collision by $\mathtt{ALICE}$ Collaboration without any free adjustable parameter~(see Table \ref{tab:Veta15}). We also show that our model based on AGK cutting rules performs consistently   better than the KL model in describing the experimental data in the wide range of energies.

The main results of this study are given in the attached $\mathtt{Mathematica}$ file.

\section{Acknowledgements}\label{}

We are indebted to Sergey Bondarenko   for inspiring discussions on the topic. This research is supported in part by the Council of Higher Education of Israel through the grant "Support program of the High Energy Physics".

\end{document}